# Performance Evaluation for *MIMO In Vivo* WBAN System*s*

Chao He, Yang Liu, Thomas P. Ketterl, Gabriel E. Arrobo, and Richard D. Gitlin

Department of Electrical Engineering
University of South Florida, Tampa, Florida 33620, USA
Email: {chaohe, yangl}@mail.usf.edu, ketterl@usf.edu, garrobo@mail.usf.edu, richgitlin@usf.edu

*Abstract*— In this paper we present the performance evaluation for a *MIMO in vivo* WBAN system, using ANSYS HFSS and the associated complete Human Body Model. We analyzed MIMO system capacity statistically and FER performance based upon an IEEE 802.11n system model, with receiver antennas placed at various angular positions around the human body. We also analyzed MIMO system capacity with receiver antennas at the front of the body at various distances from transmitter antennas. The results were compared to SISO arrangements and we demonstrate that by using 2x2 *MIMO in vivo*, better performance can be achieved, and significantly higher system capacity can be achieved when receiver antennas are located at the back of the body and in front of the body.

## I. INTRODUCTION

One appealing aspect of the emerging *Internet of Things* is to consider *in vivo* networking for Wireless Body Area Networks [WBANs] as a rich application domain for wireless technology in facilitating continuous wirelessly enabled healthcare. Due to the lossy nature of the *in vivo* medium, achieving high data rates with reliable performance will be a challenge, especially since the *in vivo* antenna performance may be affected by near-field coupling to the lossy medium and the signals levels will be limited by specified the Specific Absorption Rate (SAR) levels. SAR is the specific absorption rate of power absorption by human organs and is limited by the FCC, which in turn limits the transmission power. One potential application for *MIMO in vivo* communications is the *MARVEL* (Miniature Anchored Remote Videoscope for Expedited Laparoscopy), which is a wireless research platform for advancing MIS (Minimally Invasive Surgery), that requires high bit rates (~80–100 Mbps) for high-definition transmission and low latency during surgery [1].

In [2], the Bit Error Rate (BER) for a *MIMO in vivo* system was first analyzed. The results were compared to *SISO in vivo* and it was demonstrated that by using 2x2 *MIMO in vivo*, significant performance gains can be achieved with maximum SAR levels met [3], making it possible to achieve target data rates of 100 Mbps if the distance between Tx and Rx antennas is within 9.5 cm. To better support practical WBAN systems, the capacity and Frame Error Rate (FER) performance for *MIMO in vivo* with a distance between transmit and receive antennas greater than 9.5 cm at 2.4 GHz band and at various angular positions of the receiver around the human body, i.e. front, right side, left side, back, are analyzed in this paper. The *MIMO in vivo* system capacity is the upper limit that can be achieved in practical systems, which provides guidance on how to optimize the *MIMO in vivo* system.

This paper is organized as follows: In section II, we present the *MIMO in vivo* capacity formulas based upon IEEE 802.11n system. Section III and IV present the evaluation methods and results for *MIMO in vivo*, respectively. Finally, in section V, we present our conclusions.

## II. *MIMO IN VIVO* CAPACITY

### A. MIMO In Vivo Capacity [4]

Assuming two transmitter and receiver antennas are used in the *MIMO in vivo* system. The system can be modeled as:
$$Y_k = H_k X_k + W_k, \quad (1)$$
where $Y_k, X_k, W_k \in \mathbb{C}^2$ denote the received signal, transmitted signal, and white Gaussian noise with power density of $N_0$ respectively at subcarrier k. $H_k \in \mathbb{C}^{2*2}$ denotes the complex frequency channel response matrix at subcarrier k.

The SVD (singular value decomposition) of $H_k$ is given as:
$$H_k = U_k \Lambda_k V_k, \quad (2)$$
where $U_k, V_k \in \mathbb{C}^{2*2}$ are unitary matrices, and $\Lambda_k$ is the nonnegative diagonal matrix whose diagonal elements are singular values of $\sqrt{\lambda_{k1}}, \sqrt{\lambda_{k2}}$ respectively.

The system capacity for subcarrier k is:
$$C_k = \sum_{i=1}^{2} \log(1 + \frac{\lambda_{k1} P}{2N_0 \cdot BW}) \text{ bits/OFDM symbol}, \quad (3)$$
where P is the total transmit signal power, and BW is the configured system bandwidth in Hz. The total system capacity is calculated as:
$$C = \frac{1}{BW \cdot T_{sym}} \sum_{k=1}^{N_{data}} C_k \text{ bits/s/Hz}, \quad (4)$$
where $N_{data}$ is the total number of subcarriers configured in the system to carry data and $T_{sym}$ is the duration of each OFDM symbol.

### B. SISO In Vivo Capacity

For a performance comparison with *MIMO in vivo*, the *SISO in vivo* capacity is also calculated. The SISO system model is the same as defined in (1) except $Y_k, X_k, W_k \in \mathbb{C}^1$.

TABLE I
LOCATIONS OF ANTENNAS WITH RESPECT TO THE ORIGIN (X=0, Y=0) SHOWN IN FIG. 1

| Cases | MIMO | | | | SISO | | | | Notes |
|---|---|---|---|---|---|---|---|---|---|
| | Receiver Antennas | | Transmitter Antennas | | Receiver Antenna | | Transmitter Antenna | | |
| | X (mm) | Y (mm) | X (mm) | Y (mm) | X (mm) | Y (mm) | X (mm) | Y (mm) | |
| 1 | 300 | ±50 | 0 | ±14 | 300 | 0 | 0 | 0 | Front of body |
| 2 | ±50 | 300 | ±14 | 0 | 0 | 300 | 0 | 0 | Right side of body |
| 3 | ±50 | -300 | ±14 | 0 | 0 | -300 | 0 | 0 | Left side of body |
| 4 | -300 | ±50 | 0 | ±14 | -300 | 0 | 0 | 0 | Back of body |
| 5 | 200 | ±50 | 0 | ±14 | 200 | 0 | 0 | 0 | Front of body |
| 6 | 130 | ±50 | 0 | ±14 | 130 | 0 | 0 | 0 | Front of body |
| 7 | 100 | ±50 | 0 | ±14 | 100 | 0 | 0 | 0 | Front of body |
| 8 | 70 | ±50 | 0 | ±14 | 70 | 0 | 0 | 0 | Front of body |

Therefore, the system capacity for *SISO in vivo* is:

$$C = \frac{1}{BW \cdot T_{sym}} \sum_{k=1}^{N_{data}} \log(1 + \frac{H_k P}{N_0 \cdot BW}) \quad \text{bits/s/Hz}, \quad (5)$$

where $H_k \in \mathbb{C}^1$, P and $N_{data}$ are the same as for *MIMO in vivo*.

## III. EVALUATION METHODS

### A. Human Body Model

The simulations for the electromagnetic wave propagation were performed in ANSYS HFSS 15.0.3 using the ANSYS Human Body Model. The antennas used in the simulations were monopoles designed to operate at the 2.4 GHz band [5].

As shown in Fig.1, two transmit antennas (Tx) are placed inside the abdomen while two receive antennas (Rx) are placed at different locations around the body at the same planar height, as given in Table I. Cases 1-4 are cases with the same distance between Tx and Rx antennas, but with different angular positions, which correspond to the front, right side, left side, and back body, respectively. Cases 1, 5-8 are cases with the Rx antennas in front of the body with the same angular positions, but with varying distances between Tx and Rx antennas. It should be noted, since the permittivity of the body is much higher than that of free space, the wavelength is smaller inside the body and varies as it passes through various tissues and organs. On average, the wavelength is the reduced by the square root of the dielectric constant and is approximately six times smaller *in vivo* than in free space.

### B. Evaluation Methods

The system capacity analysis and FER performance in the *in vivo* environment have been performed based on the IEEE 802.11n standard [6] transceiver. Agilent SystemVue is used to simulate the FER performance. Because of the form factor constraint inside the human body, our initial study is restricted to 2x2 MIMO. The system bandwidth of 20 MHz is used in the evaluation. The 802.11n standard supports different Modulation and Coding Schemes (MCS) represented by a MCS index. The transmission power is set to be 0.412 mw [3], which gives the maximum local SAR level of 1.48 W/kg that will not exceed the maximum allowable SAR level of 1.6 W/kg. The thermal noise power is set to -101 dBm. Hence, in the system capacity analysis, the parameters in (3)-(5) are determined as follows:

$P = 0.412$ mw, $N_0 = -174$ dBm, $BW = 20$ MHz, $N_{data} = 52$, $T_{sym} = 4$ us.

## IV. EVALUATION RESULTS

The system capacity for both *MIMO* and *SISO in vivo* can be calculated based upon (2)-(5). The FER for the IEEE 802.11n system was found by transmitting 100,000 frames for each simulation for different MCS index values.

Figure 2 shows the system capacity for different angular positions around the human body with the same distance between Tx and Rx antennas of 300 mm. From Fig. 2, we can observe the capacity gain compared with corresponding SISO cases, where the greatest capacity gain can be seen in the case of both MIMO antennas at the back of the body. We can also see from Fig. 2 that the system capacity of *MIMO in vivo* for the cases of front and back body is much better than that of the other two cases of side body antennas. This is because much higher attenuation exists inside the body due to the greater *in vivo* distance for the two cases of side body. This is also verified by the FER performance result in Fig. 3. Furthermore, compared with the other three cases, *MIMO in vivo* for the back body case performs the best. From Fig. 2, we see that with the greater distance between Tx and Rx antennas, the system capacity will fall below 1.4 bits/s/Hz for whatever angular positions the receiver antennas are located. Hence, for a 20 MHz system bandwidth, only a data rate of less than 28 Mbps can be supported, which is a motivation for us to use a relay or other forms of cooperative networked communications and/or place the receiver antennas as close to, or on, the front or back of the body to support a data rate as high as 100 Mbps. Of course, increasing the bandwidth to 40 MHz will double the achievable bit rate.

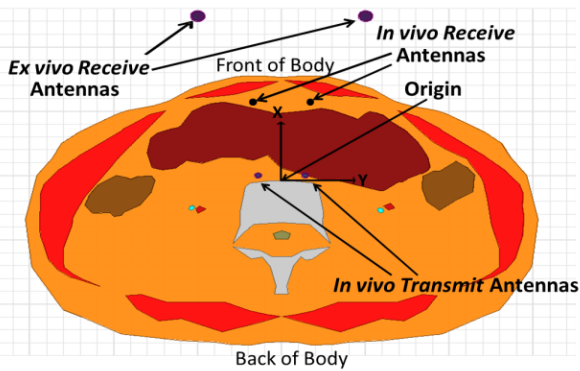

Fig. 1. Simulation setup showing locations of the MIMO antennas.

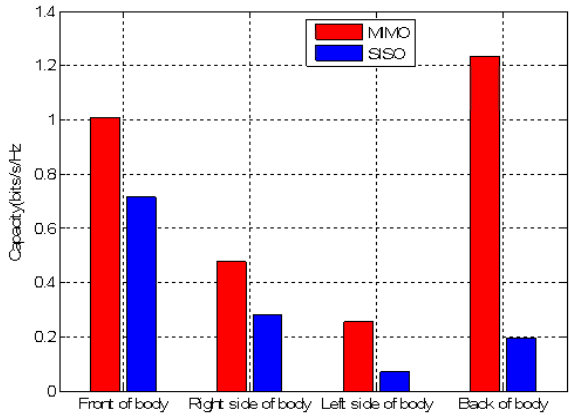

Fig. 2. (2x2) *MIMO* and *SISO in vivo* system capacity comparison for front, right side, left side, and back of the body.

Figure 3 shows the FER performance for varying angular positions but with the same distance of 300 mm between the Tx and Rx antennas. From Fig. 3, we can find that much lower FER performance can be achieved for the back and front body cases where the case for back body performs the best, which is consistent with the analysis of MIMO system capacity in Fig. 2.

Figure 4 shows the system capacity for the cases of Rx

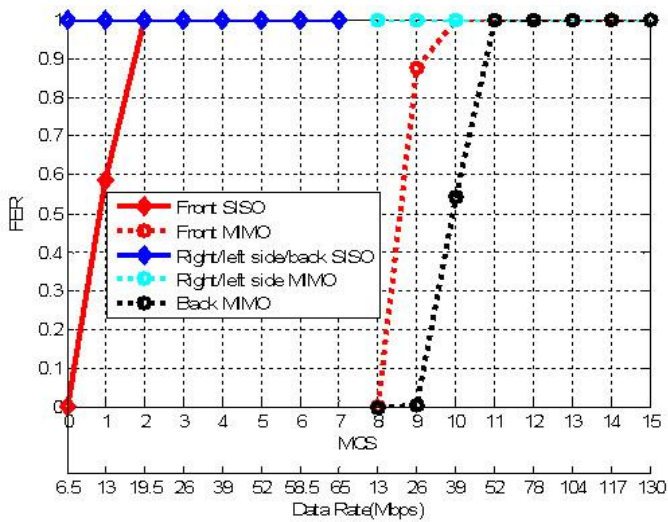

Fig. 3. (2x2) *MIMO* and *SISO in vivo* FER performance comparison for front, right side, left side, and back of the body.

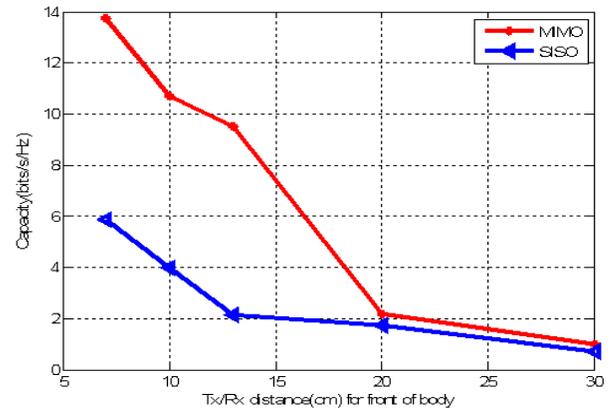

Fig. 4. (2x2) *MIMO* and *SISO in vivo* capacity comparison as function of the distance of the Tx and Rx antennas in front of the antennas in front of the body with varying distances between Tx and Rx antennas. It can be seen that much less capacity can be achieved with increasing distance. To support the required data rate of 100 Mbps, the capacity cannot be less than 5 bit/s/Hz (i.e. 100 Mbps/20 MHz), thus the distance cannot be greater than ~17cm. The system capacity decreases rapidly when the distance becomes greater, making relay or other forms of cooperative networked communications necessary in the WBAN network architecture.

## V. CONCLUSIONS

In this paper, we analyzed the system capacity and simulated the FER performance of a 2x2 *MIMO in Vivo* transceiver based upon the IEEE 802.11n standard. From the evaluation result in this study, *MIMO in vivo* can achieve better system capacity than *SISO in vivo*. Significantly better system capacity can be observed when receiver antennas are paced at the back or the front of body than when placed at the side of the body. It is also found that to meet higher data rate requirements as high as 100 Mbps with a distance between Tx and Rx antennas greater than 17 cm, relay or other similar cooperative networked communications are necessary to be introduced into the WBAN network.

ACKNOWLEDGEMENT

This research was supported in part by NSF Grant IIP-1217306 and the Florida 21st Century Scholars program.